\documentclass[prc,aps,twocolumn,showpacs,amssymb,superscriptaddress,float]{revtex4}

\usepackage{amsmath}
\usepackage{graphicx}
\usepackage{dcolumn}
\usepackage{float}

\begin{document}

\title{Study of the ground-state energy of $^{40}$Ca with the CD-Bonn
  nucleon-nucleon potential}
\author{L. Coraggio}
\affiliation{Dipartimento di Scienze Fisiche, Universit\`a
di Napoli Federico II, \\ and Istituto Nazionale di Fisica Nucleare, \\
Complesso Universitario di Monte  S. Angelo, Via Cintia - I-80126 Napoli,
Italy}
\author{A. Covello}
\affiliation{Dipartimento di Scienze Fisiche, Universit\`a
di Napoli Federico II, \\ and Istituto Nazionale di Fisica Nucleare, \\
Complesso Universitario di Monte  S. Angelo, Via Cintia - I-80126 Napoli,
Italy}
\author{A. Gargano}
\affiliation{Dipartimento di Scienze Fisiche, Universit\`a
di Napoli Federico II, \\ and Istituto Nazionale di Fisica Nucleare, \\
Complesso Universitario di Monte  S. Angelo, Via Cintia - I-80126 Napoli,
Italy}
\author{N. Itaco}
\affiliation{Dipartimento di Scienze Fisiche, Universit\`a
di Napoli Federico II, \\ and Istituto Nazionale di Fisica Nucleare, \\
Complesso Universitario di Monte  S. Angelo, Via Cintia - I-80126 Napoli,
Italy}
\author{T. T. S. Kuo}
\affiliation{Department of Physics, SUNY, Stony Brook, New York 11794}

\date{\today}

\begin{abstract}
We have calculated the ground-state energy of the doubly-magic nucleus
$^{40}$Ca within the framework of the Goldstone expansion using the
CD-Bonn nucleon-nucleon potential. 
The short-range repulsion of this potential has been renormalized by
integrating out its high-momentum components so as to derive a
low-momentum potential $V_{\rm low-k}$ defined up to a cutoff momentum
$\Lambda$. 
A simple criterion has been employed to establish a connection between
this cutoff momentum and the size of the two-nucleon model space in
the harmonic oscillator basis.
This model-space truncation approach provides a reliable way to
renormalize the free nucleon-nucleon potential preserving its
many-body physics.
The role of the $3p-3h$ and $4p-4h$ excitations in the description of
the ground state of $^{40}$Ca is discussed.
\end{abstract}

\pacs{21.30.Fe, 21.60.Jz, 21.10.Dr,27.40.+z}

\maketitle

A fundamental goal of nuclear physics is to describe the properties of
nuclei starting from the forces among nucleons.
To this end, one has to employ many-body methods well suited to handle
the strong short-range correlations that are induced in nuclei by the
free-space nucleon-nucleon ($NN$) potential $V_{NN}$.  
In other words, the method employed should produce results which are
only slightly affected by the approximations involved, in the sense
that it should be possible to keep the latter under control by way of
convergence checks.
This first-principle approach to the description of nuclear structure
is nowadays referred to as {\it ab initio} approach \cite{Barrett05}.

In the last decade, thanks also to the considerable increase in
computer power, a substantial progress has been achieved in microscopic
approaches to the nuclear many-body problem, such as the Green's
function Monte Carlo (GFMC) \cite{Pieper01}, no-core shell model
(NCSM) \cite{Navratil00a,Navratil00b}, and coupled-cluster methods
(CCM) \cite{Suzuki94b,Dean04b}.

Historically, the first calculations on light $p$-shell nuclei
using the GFMC method were performed in the mid-1990s by
the Argonne group \cite{Pieper01}.
They employed the high-precision $NN$ potential $AV18$
\cite{Wiringa95} plus a three-body interaction, the latter being
fitted to reproduce the binding energy of selected few-body systems. 
Since then, these calculations have been successfully performed up to
mass $A=12$ \cite{Pieper05,Wiringa06}, which is the present limit of
GFMC with the available computer technology, owing to the exponential
growth of the spin-isospin vector size \cite{Barrett03}. 
This limit may be overcome by introducing effective interactions.
In this way, the NCSM and CCM allow to perform calculations beyond the
$p$-shell mass region.
More precisely, the NCSM has been applied
\cite{Navratil00a,Navratil00b} to nuclei with mass $A \leq 16$, using
either coordinate-space or momentum dependent $V_{NN}$'s
\cite{Vary05}. 
In some cases, also three-body interactions have been included
\cite{Hayes03,Navratil03,Nogga06}. 
The CCM can potentially be used for much heavier systems; in fact, in
the late 1970s it was applied to the doubly-closed nuclei $^{16}$O and 
$^{40}$Ca \cite{Kummel78}.
Actually, coupled-cluster calculations employing modern $V_{NN}$'s
\cite{Dean04b} have been recently performed for valence systems around
$^{16}$O \cite{Kowalski04,Wloch05,Gour06}. 
CCM-like calculations are also those performed for nuclei up to
$^{16}$O in Refs. \cite{Fujii04b,Fujii05}, where different
realistic $NN$ potentials have been used in the framework of the
unitary model-operator approach (UMOA)
\cite{Suzuki94b,Fujii00,Fujii04a}. 

In Refs. \cite{Bogner01,Bogner02} a method to renormalize the bare $NN$ 
interaction has been introduced, which may be considered an 
advantageous alternative to the use of the Brueckner $G$ matrix.
A low-momentum model space is defined up to a cutoff momentum
$\Lambda$, and an effective low-momentum potential $V_{\rm
  low-k}$, which satisfies the decoupling condition between the low-
and high-momentum spaces, is derived from the original $V_{NN}$. 
The $V_{\rm low-k}$ is a smooth potential which preserves exactly the
low-momentum on-shell properties of the original $V_{NN}$ and can be
used directly in nuclear structure calculations. 

Recently, we have investigated \cite{Coraggio06b,Coraggio07a} how
$\Lambda$ is related to the dimension of the configuration space in
the coordinate representation where our calculations are performed. 
We have introduced a simple criterion to map out the model space made
up by the two-nucleon states in the harmonic-oscillator (HO) basis 
according to the value of the cutoff momentum $\Lambda$.
The validity of this procedure was tested by calculating, in the
framework of the Goldstone expansion, the ground-state (g.s.) energy of
$^4$He with the CD-Bonn \cite{Machleidt01b}, N$^3$LO \cite{Entem03},
and Bonn A \cite{Machleidt87} potentials, and comparing
the results with those obtained using the Faddeev-Yakubovsky method.
Taking into account perturbative contributions up to fourth order in 
$V_{\rm low-k}$, we have found that the energy differences are at most
390 keV.  
The limited size of the discrepancies shows that this approach
provides a reliable way to renormalize the $NN$ potential preserving
the physics beyond the two-body system too.
 
We also performed calculations for heavier systems, such as
$^{16}$O and $^{40}$Ca, and obtained converged results for the CD-Bonn
$NN$ potential using a limited number of oscillator quanta. 
As regards $^{40}$Ca, the g.s. energy was calculated including
Goldstone diagrams only up to third order in $V_{\rm low-k}$. 

It seems fair to say that, at present, an {\it ab initio}
calculation for $^{40}$Ca represents a major step on the way to the
fully microscopic description of nuclear systems beyond $^{16}$O.
In this work, we improve our calculation of the $^{40}$Ca g.s. energy
including all the fourth-order contributions.
A main motivation for this extension of our calculation of the
g.s. energy of $^{40}$Ca is to study the role of a higher-order class
of excitations, namely the $3p-3h$ and $4p-4h$ ones, which come into
play starting from the fourth order of the Goldstone expansion.
It should be pointed out that this is the first fully microscopic
study of this nucleus, apart from a very preliminary calculation by
Kumagai {\it et al.} \cite{Kumagai97} and Fujii {\it et al.}
\cite{Fujii05} in the framework of the unitary model-operator
approach. 

The first step of our calculation is to renormalize the short-range
repulsion of the $NN$ potential by integrating out its high momentum
components through the so-called $V_{\rm low-k}$ approach (see
Refs. \cite{Bogner02,Bogner03}). 
The low-momentum potential $V_{\rm low-k}$ preserves the physics of
the two-nucleon system, and consequently the $\chi^2/{datum}$ of the
original $NN$ potential, up to the cutoff momentum $\Lambda$. 
As mentioned above, it is a smooth potential and can therefore be used
directly in a perturbative nuclear structure calculation. 

The $V_{\rm low-k}$ is defined in the momentum space, and it
is desirable to relate the cutoff momentum $\Lambda$ to the size of
the harmonic-oscillator (HO) space in the coordinate representation
\cite{Coraggio06b,Coraggio07a}, where we perform our calculations for
finite nuclei. 

Let us consider the relative motion of two nucleons in a HO well in
the momentum representation. 
For a given maximum relative momentum $\Lambda$, the corresponding
maximum value of the energy is:

\begin{equation}
E_{\rm max} = \frac{ \hbar^2 \Lambda^2}{M}~~,
\label{one}
\end{equation}

\noindent
where $M$ is the nucleon mass.

This relation may be rewritten in terms of the maximum number 
$N_{\rm max}$ of HO quanta in the relative coordinate system. 
For a given HO parameter $\hbar \omega$ we have:

\begin{equation}
\left( N_{\rm max} + \frac{3}{2} \right) \hbar \omega = \frac{ 
\hbar^2 \Lambda^2}{M}~~.
\label{two}
\end{equation}

\noindent
The above equation provides a simple criterion to map out the
two-nucleon HO model space.
If we write the two-nucleon states as the product of HO wave functions

\begin{equation}
|a~b \rangle = | n_a l_a j_a,~n_b l_b j_b \rangle~~,
\label{three}
\end{equation}

\noindent
our HO model space is defined as spanned by those two-nucleon states that
satisfy the constraint

\begin{equation}
2n_a+l_a+2n_b+l_b \leq N_{\rm max}~~.
\label{four}
\end{equation}

Making use of the above approach, in this paper we have studied the
g.s. energy of $^{40}$Ca within the framework of the
Goldstone expansion \cite{Goldstone57}. 
We start from the purely intrinsic hamiltonian 

\begin{equation}
H= \left( 1 - \frac{1}{A} \right) \sum_{i=1}^{A} \frac{p_i^2}{2M} + \sum_{i<j}
\left( V_{ij} - \frac{ {\rm {\bf p}}_i \cdot {\rm {\bf p}}_j }{MA} \right)~~, 
\end{equation}

\noindent
where $V_{ij}$ stands for the renormalized $V_{NN}$ potential plus the 
Coulomb force, and construct the Hartree-Fock (HF) basis expanding the
HF single particle (SP) states in terms of HO wave functions.
The next step is to sum up all the Goldstone linked diagrams for the
ground-state energy up to fourth-order in the two-body interaction.
The complete list of the fourth-order diagrams can be found in
Ref. \cite{Ayoub79,Bartlett81}. 
Using Pad\'e approximants \cite{Baker70,Ayoub79} one may obtain a value
to which the perturbation series should converge.
In this work, we report results obtained using the Pad\'e approximant
$[2|2]$, whose explicit expression is

\begin{equation}
[ 2|2 ]=
\frac{E_0(1+\gamma_1+\gamma_2)+E_1(1+\gamma_2)+E_2}{1+\gamma_1+
\gamma_2}~~, \label{pade}
\end{equation}

\noindent
where 

\[
\gamma_1 = \frac{E_2E_4-E_3^2}{E_1E_3-E_2^2} ~~,~~~~~
\gamma_2 = -\frac{E_3+E_1 \gamma_1}{E_2} \nonumber ~~,
\]

\noindent
$E_i$ being the $i$th order energy contribution in the Goldstone
expansion.

In principle, our results should not depend on the HO parameter $\hbar 
\omega$, whose value characterizes the HO wavefunctions employed to
expand the HF SP states.
Actually, our calculations are made in a truncated model space, whose size 
is related to the values of the cutoff momentum $\Lambda$ and the $\hbar 
\omega$ parameter by relations (\ref{two}) and (\ref{four}).
Obviously, for $N_{\rm max} \rightarrow \infty$ this dependence
disappears.
So, we perform our calculations increasing the $N_{\rm max}$ value
(and consequently $\Lambda$) for different $\hbar \omega$ values.
Finally, we choose the results which correspond to the value of $\hbar
\omega$ for which they are practically independent of $N_{\rm max}$.

\begin{figure}[H]
\begin{center}
\includegraphics[scale=0.55,angle=0]{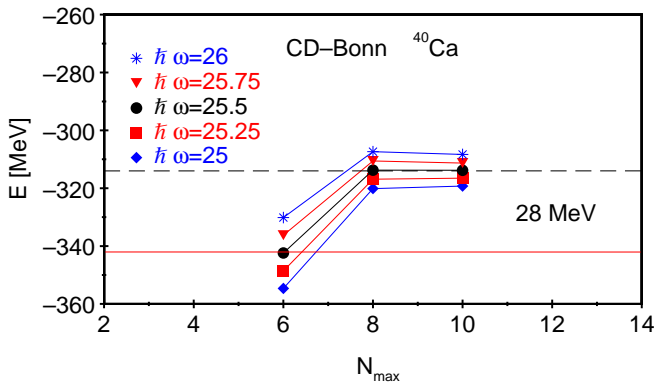}
\caption{(Color online) Ground-state energy of $^{40}$Ca with the CD-Bonn 
potential as a function of $N_{\rm max}$, for different values of 
$\hbar \omega$. 
The straight line represents the experimental value, while the dashed
one our converged result.
The value of the energy difference between the former and the latter
is also reported.}
\label{40cacdbonn}
\end{center}
\end{figure}

In Fig. \ref{40cacdbonn} we report the ground-state energy of $^{40}$Ca 
calculated with the CD-Bonn potential \cite{Machleidt01b}. 
The straight red line indicates the experimental datum \cite{Audi03}
while the other curves represent our calculated values, for different
values of $\hbar \omega$, versus the maximum number of HO quanta
$N_{\rm max}$ that limits the two-nucleon configurations according to
relation (\ref{four}).

We obtain convergence for $\hbar \omega=$ 25.5 MeV, and the
corresponding energy is $(-314 \pm 3)$ MeV, as indicated in
Fig. \ref{40cacdbonn} by the dashed line.
The error has been evaluated taking into account the dependence of our
results on $\hbar \omega$ \cite{Coraggio07a}.
In Ref. \cite{Coraggio06b} we calculated the $^{40}$Ca g.s.
energy taking into account all the contributions in the
Goldstone expansion up to third order.
The converged value, obtained with the Pad\`e approximant $[2|1]$
turned out to be ($-308 \pm 3$) MeV with $\hbar \omega=$ 25.5 MeV.
The difference between the two results is only about $2\%$ of the
total binding energy, the converged value at fourth order being 6 MeV
more attractive than the third-order one.
This is in line with the outcome of our calculations 
\cite{Coraggio06b,Coraggio07a} of the g.s. energy of $^4$He and
$^{16}$O with the CD-Bonn potential. 
In those cases the converged fourth-order result was 0.6 MeV and 3 MeV
more attractive, respectively, than the converged third-order one.

\begin{table}[H]
\caption{Calculated $2p-2h$, $3p-3h$, and $4p-4h$ fourth-order 
contributions (in MeV) to the g.s. energy of $^{40}$Ca with the
CD-Bonn potential.
Calculations correspond to $N_{\rm max}=10$.} 
\begin{ruledtabular}
\begin{tabular}{lcccc}
\colrule
 Nucleus & $2p-2h$ & $3p-3h$ & $4p-4h$ & 4th order\\
 $^{40}$Ca   & -15 & -16 & +24 & -7 \\
\end{tabular}
\end{ruledtabular}
\label{table1}
\end{table}

It is worth now to make a brief discussion about the role of $3p-3h$
and $4p-4h$ contributions which, as pointed out before, come into
play only at the fourth order of the Goldstone expansion and
beyond. 
To this end, we report in Table \ref{table1} the $2p-2h$, $3p-3h$, and
$4p-4h$ fourth-order contributions to the g.s. energy of $^{40}$Ca
with the CD-Bonn potential. 
Since we use the Hartree-Fock basis, all $1p-1h$ excitations of the
ground state are identically zero. 
The inspection of Table \ref{table1} evidences that the role played by
the $3p-3h$ and $4p-4h$ excitations is significant, their net
repulsive contribution being 8 MeV.
The latter counterbalances the fourth-order contribution of the
$2p-2h$ excitations, so that the total fourth-order contribution is
only -7 MeV.

As regards the comparison with experiment, our calculated $^{40}$Ca
binding energy underestimates the experimental one by  28 MeV, which
is $8\%$ of the experimental binding energy. 
Our calculations of $^4$He and $^{16}$O with the CD-Bonn potential
\cite{Coraggio06b,Coraggio07a} show the same percent difference 
from the experimental data.
This seems to confirm the need of a three-body force in addition to
the $NN$ CD-Bonn potential, in order to compensate for the lack of
attraction of the latter.

In conclusion, this work presents a fully microscopic calculation of
the ground-state energy of $^{40}$Ca with the CD-Bonn $NN$ potential.
We hope that this may stimulate, and provide some useful hints to,
future non-perturbative {\it ab initio} calculations.

\begin{acknowledgments}
This work was supported in part by the Italian Ministero
dell'Istruzione, dell'Universit\`a e della Ricerca  (MIUR), and by the
U.S. DOE Grant No. DE-FG02-88ER40388. 
\end{acknowledgments}

\bibliographystyle{apsrev}
\bibliography{biblio}

\end{document}